\documentclass[english,prl,aps,twocolumn,floatfix,showpacs,tightenlines,superscriptaddress]{revtex4}
\usepackage{amsmath,amsfonts,amssymb,latexsym,times}
\usepackage{verbatim}
\usepackage{graphicx}
\usepackage{bm}
\usepackage{epstopdf}
\usepackage[dvipsnames]{xcolor} 
\usepackage{pstricks,pst-fill,pst-text,pst-node,pstricks-add}

\newcommand{\be}{\begin{equation}}
\newcommand{\ee}{\end{equation}}
\newcommand{\ba}{\begin{eqnarray}}
\newcommand{\ea}{\end{eqnarray}}

\begin{document}
\title{Universality classes of dense polymers and conformal sigma models}
\author{C. Candu}
\affiliation{Institut de Physique Th\'eorique CEA, IPhT, CNRS, URA 2306, F-91191 Gif-sur-Yvette, France }
\author{J.L. Jacobsen}
\affiliation{Laboratoire de Physique Th\'eorique, Ecole Normale Sup\'erieure, 24 rue Lhomond, 75005 Paris, France}
\affiliation{Institut de Physique Th\'eorique CEA, IPhT, CNRS, URA 2306, F-91191 Gif-sur-Yvette, France }
\author{N. Read}
\affiliation{Department of Physics, Yale University, P.O. Box 208120, New Haven,
CT 06520-8120, USA}
\author{H. Saleur}
\affiliation{Institut de Physique Th\'eorique CEA, IPhT, CNRS, URA 2306, F-91191 Gif-sur-Yvette, France }
\affiliation{Department of Physics and Astronomy, University of Southern California, Los Angeles, CA 90089-0484}

\date{\today}

\begin{abstract}
In the usual statistical model of a dense polymer (a single
space-filling loop on a lattice) in two dimensions the loop does not
cross itself. We modify this by including intersections in which {\em
three} lines can cross at the same point, with some statistical weight
$w$ per crossing. We show that our model describes a line of critical
theories with continuously-varying exponents depending on $w$, described
by a conformally-invariant non-linear sigma model with varying coupling
constant $g_\sigma^2\geq 0$. For the boundary critical behavior, or the
model defined in a strip, we propose an exact formula for the $\ell$-leg
exponents,
$h_\ell=g_\sigma^2 \ell(\ell-2)/8$, which is shown numerically to hold
very well.
\end{abstract}


\maketitle

Loop models are ubiquitous in low dimensional statistical mechanics, and have been studied for decades  \cite{Nienhuis}. They have recently grown to play a major role in  topological quantum computing \cite{Freedman}. 

Most loop models studied so far  have to forbid intersections to be solvable. Their critical exponents can then be calculated using techniques of conformal field theory \cite{CG}, Coulomb gas, or 
stochastic Loewner evolution (SLE)  \cite{SLE,Werner}.



Self avoiding walks, whose long distance properties describe real polymers at interfaces, 
are the simplest of all loop models. 
In the ordinary,  so-called {\em dilute} case, it is known that allowing intersections (hence obtaining
self-avoiding trails) does not change the long-distance properties
\cite{Guttman}. The  {\em dense} case, where a single self avoiding loop on a lattice  is forced to occupy a finite fraction of the sites (and thus resembles a  real polymer in a melt), is different.  Allowing intersections  does take the model to another
 universality class \cite{JRS}, which, however, shares many
features with ordinary Brownian motion, and does not seem to exhibit new
families of critical exponents. The first important result of this paper is that only  allowing intersections where {\em three} lines cross simultaneously produces a  very different behavior: a line of critical points is obtained, with central charge $c=-2$, and  continuously varying critical exponents,

Our  second important result concerns the nature of this critical line, which is very unusual in statistical mechanics.  

A  convenient way to describe many loop models  is to use a supersymmetric  (SUSY) formulation,  in which  the degrees of freedom can take bosonic or fermionic values \cite{ParisiSourlas}, and the action is invariant under the action of a supergroup. The resulting field theories are however  difficult to solve, in part due the lack of unitarity, and of current algebra symmetry. 

In the last few years, progress on one kind of such theories -  $\sigma$-models on supergroups or supercosets - has been achieved in the framework of the AdS/CFT conjecture \cite{berk,bersh}.  An archetypal example is the principal chiral (PCM)
model on  $PSL(2|2)$, which was found  to be massless
for a large range of values of the coupling constant
$g_\sigma^2$. This is very different from what happens in ordinary groups, such as $SU(2)$, where  the
PCM exhibits asymptotic freedom and spontaneous mass generation. The presence of conformal invariance and (super)group symmetry is very interesting and potentially useful, yet, despite a lot of work, no complete solution of even the $PSL(2|2)$ case, has been achieved \cite{GQS}. 

We show in this paper  that allowing crossings in dense polymers leads
to close cousins of the $PSL(n|n)$ models: $\sigma$-models on
superprojective spaces (the super-analogs of ordinary projective spaces) $U(n|n)/U(n-1|n)\times U(1)$. This
identification has crucial consequences. It bridges the study of loop models with the one of $\sigma$ models, it gives direct access to properties
of the $\sigma$-models both numerically, and, potentially,
analytically using the techniques developed in \cite{RS2,RS3}. It also
allows the determination of the critical exponents in the original geometrical problem. 

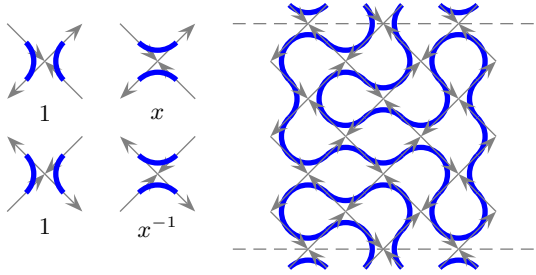
\begin{figure}
$$
\begin{pspicture}(-3.0,0.2)(4.0,3.1)
 \psellipticarc[linecolor=blue,linewidth=2.0pt]{-}(1.0,-0.5)(0.353553,0.353553){45}{135}
 \psellipticarc[linecolor=blue,linewidth=2.0pt]{-}(3.0,-0.5)(0.353553,0.353553){45}{135}
 \psellipticarc[linecolor=blue,linewidth=2.0pt]{-}(1.5,0.0)(0.353553,0.353553){315}{135}
 \psellipticarc[linecolor=blue,linewidth=2.0pt]{-}(2.5,0.0)(0.353553,0.353553){135}{225}
 \psellipticarc[linecolor=blue,linewidth=2.0pt]{-}(1.0,0.5)(0.353553,0.353553){45}{315}
 \psellipticarc[linecolor=blue,linewidth=2.0pt]{-}(2.0,0.5)(0.353553,0.353553){315}{135}
 \psellipticarc[linecolor=blue,linewidth=2.0pt]{-}(3.0,0.5)(0.353553,0.353553){135}{45}
 \psellipticarc[linecolor=blue,linewidth=2.0pt]{-}(1.5,1.0)(0.353553,0.353553){45}{135}
 \psellipticarc[linecolor=blue,linewidth=2.0pt]{-}(1.5,1.0)(0.353553,0.353553){225}{315}
 \psellipticarc[linecolor=blue,linewidth=2.0pt]{-}(2.5,1.0)(0.353553,0.353553){315}{135}
 \psellipticarc[linecolor=blue,linewidth=2.0pt]{-}(3.5,1.0)(0.353553,0.353553){135}{225}
 \psellipticarc[linecolor=blue,linewidth=2.0pt]{-}(1.0,1.5)(0.353553,0.353553){135}{315}
 \psellipticarc[linecolor=blue,linewidth=2.0pt]{-}(2.0,1.5)(0.353553,0.353553){45}{135}
 \psellipticarc[linecolor=blue,linewidth=2.0pt]{-}(2.0,1.5)(0.353553,0.353553){225}{315}
 \psellipticarc[linecolor=blue,linewidth=2.0pt]{-}(3.0,1.5)(0.353553,0.353553){315}{45}
 \psellipticarc[linecolor=blue,linewidth=2.0pt]{-}(0.5,2.0)(0.353553,0.353553){315}{45}
 \psellipticarc[linecolor=blue,linewidth=2.0pt]{-}(1.5,2.0)(0.353553,0.353553){45}{315}
 \psellipticarc[linecolor=blue,linewidth=2.0pt]{-}(2.5,2.0)(0.353553,0.353553){225}{45}
 \psellipticarc[linecolor=blue,linewidth=2.0pt]{-}(3.5,2.0)(0.353553,0.353553){135}{225}
 \psellipticarc[linecolor=blue,linewidth=2.0pt]{-}(1.0,2.5)(0.353553,0.353553){45}{225}
 \psellipticarc[linecolor=blue,linewidth=2.0pt]{-}(2.0,2.5)(0.353553,0.353553){225}{45}
 \psellipticarc[linecolor=blue,linewidth=2.0pt]{-}(3.0,2.5)(0.353553,0.353553){315}{225}
 \psellipticarc[linecolor=blue,linewidth=2.0pt]{-}(1.5,3.0)(0.353553,0.353553){225}{45}
 \psellipticarc[linecolor=blue,linewidth=2.0pt]{-}(2.5,3.0)(0.353553,0.353553){135}{225}
 \psellipticarc[linecolor=blue,linewidth=2.0pt]{-}(1.0,3.5)(0.353553,0.353553){225}{315}
 \psellipticarc[linecolor=blue,linewidth=2.0pt]{-}(3.0,3.5)(0.353553,0.353553){225}{315}
 \psline[linecolor=gray,linewidth=0.5pt,arrowsize=5pt]{->}(1.0,1.0)(0.5,0.5)
 \psline[linecolor=gray,linewidth=0.5pt,arrowsize=5pt]{->}(1.0,2.0)(0.5,1.5)
 \psline[linecolor=gray,linewidth=0.5pt,arrowsize=5pt]{->}(1.0,3.0)(0.5,2.5)
 \psline[linecolor=gray,linewidth=0.5pt,arrowsize=5pt]{->}(2.0,1.0)(1.5,0.5)
 \psline[linecolor=gray,linewidth=0.5pt,arrowsize=5pt]{->}(2.0,2.0)(1.5,1.5)
 \psline[linecolor=gray,linewidth=0.5pt,arrowsize=5pt]{->}(2.0,3.0)(1.5,2.5)
 \psline[linecolor=gray,linewidth=0.5pt,arrowsize=5pt]{->}(3.0,1.0)(2.5,0.5)
 \psline[linecolor=gray,linewidth=0.5pt,arrowsize=5pt]{->}(3.0,2.0)(2.5,1.5)
 \psline[linecolor=gray,linewidth=0.5pt,arrowsize=5pt]{->}(3.0,3.0)(2.5,2.5)
 \psline[linecolor=gray,linewidth=0.5pt,arrowsize=5pt]{->}(0.5,0.5)(1.0,0.0)
 \psline[linecolor=gray,linewidth=0.5pt,arrowsize=5pt]{->}(0.5,1.5)(1.0,1.0)
 \psline[linecolor=gray,linewidth=0.5pt,arrowsize=5pt]{->}(0.5,2.5)(1.0,2.0)
 \psline[linecolor=gray,linewidth=0.5pt,arrowsize=5pt]{->}(1.5,0.5)(2.0,0.0)
 \psline[linecolor=gray,linewidth=0.5pt,arrowsize=5pt]{->}(1.5,1.5)(2.0,1.0)
 \psline[linecolor=gray,linewidth=0.5pt,arrowsize=5pt]{->}(1.5,2.5)(2.0,2.0)
 \psline[linecolor=gray,linewidth=0.5pt,arrowsize=5pt]{->}(2.5,0.5)(3.0,0.0)
 \psline[linecolor=gray,linewidth=0.5pt,arrowsize=5pt]{->}(2.5,1.5)(3.0,1.0)
 \psline[linecolor=gray,linewidth=0.5pt,arrowsize=5pt]{->}(2.5,2.5)(3.0,2.0)
 \psline[linecolor=gray,linewidth=0.5pt,arrowsize=5pt]{<-}(1.0,1.0)(1.5,0.5)
 \psline[linecolor=gray,linewidth=0.5pt,arrowsize=5pt]{<-}(1.0,2.0)(1.5,1.5)
 \psline[linecolor=gray,linewidth=0.5pt,arrowsize=5pt]{<-}(1.0,3.0)(1.5,2.5)
 \psline[linecolor=gray,linewidth=0.5pt,arrowsize=5pt]{<-}(2.0,1.0)(2.5,0.5)
 \psline[linecolor=gray,linewidth=0.5pt,arrowsize=5pt]{<-}(2.0,2.0)(2.5,1.5)
 \psline[linecolor=gray,linewidth=0.5pt,arrowsize=5pt]{<-}(2.0,3.0)(2.5,2.5)
 \psline[linecolor=gray,linewidth=0.5pt,arrowsize=5pt]{<-}(3.0,1.0)(3.5,0.5)
 \psline[linecolor=gray,linewidth=0.5pt,arrowsize=5pt]{<-}(3.0,2.0)(3.5,1.5)
 \psline[linecolor=gray,linewidth=0.5pt,arrowsize=5pt]{<-}(3.0,3.0)(3.5,2.5)
 \psline[linecolor=gray,linewidth=0.5pt,arrowsize=5pt]{<-}(1.5,0.5)(1.0,0.0)
 \psline[linecolor=gray,linewidth=0.5pt,arrowsize=5pt]{<-}(1.5,1.5)(1.0,1.0)
 \psline[linecolor=gray,linewidth=0.5pt,arrowsize=5pt]{<-}(1.5,2.5)(1.0,2.0)
 \psline[linecolor=gray,linewidth=0.5pt,arrowsize=5pt]{<-}(2.5,0.5)(2.0,0.0)
 \psline[linecolor=gray,linewidth=0.5pt,arrowsize=5pt]{<-}(2.5,1.5)(2.0,1.0)
 \psline[linecolor=gray,linewidth=0.5pt,arrowsize=5pt]{<-}(2.5,2.5)(2.0,2.0)
 \psline[linecolor=gray,linewidth=0.5pt,arrowsize=5pt]{<-}(3.5,0.5)(3.0,0.0)
 \psline[linecolor=gray,linewidth=0.5pt,arrowsize=5pt]{<-}(3.5,1.5)(3.0,1.0)
 \psline[linecolor=gray,linewidth=0.5pt,arrowsize=5pt]{<-}(3.5,2.5)(3.0,2.0)
 \psline[linecolor=gray,linewidth=0.5pt,arrowsize=5pt]{->}(0.75,3.25)(1.0,3.0)
 \psline[linecolor=gray,linewidth=0.5pt,arrowsize=5pt]{-}(1.25,3.25)(1.0,3.0)
 \psline[linecolor=gray,linewidth=0.5pt,arrowsize=5pt]{->}(1.75,3.25)(2.0,3.0)
 \psline[linecolor=gray,linewidth=0.5pt,arrowsize=5pt]{-}(2.25,3.25)(2.0,3.0)
 \psline[linecolor=gray,linewidth=0.5pt,arrowsize=5pt]{->}(2.75,3.25)(3.0,3.0)
 \psline[linecolor=gray,linewidth=0.5pt,arrowsize=5pt]{-}(3.25,3.25)(3.0,3.0)
 \psline[linecolor=gray,linewidth=0.5pt,arrowsize=5pt]{-}(0.75,-0.25)(1.0,0.0)
 \psline[linecolor=gray,linewidth=0.5pt,arrowsize=5pt]{->}(1.25,-0.25)(1.0,0.0)
 \psline[linecolor=gray,linewidth=0.5pt,arrowsize=5pt]{-}(1.75,-0.25)(2.0,0.0)
 \psline[linecolor=gray,linewidth=0.5pt,arrowsize=5pt]{->}(2.25,-0.25)(2.0,0.0)
 \psline[linecolor=gray,linewidth=0.5pt,arrowsize=5pt]{-}(2.75,-0.25)(3.0,0.0)
 \psline[linecolor=gray,linewidth=0.5pt,arrowsize=5pt]{->}(3.25,-0.25)(3.0,0.0)
 \psline[linecolor=gray,linewidth=0.5pt,linestyle=dashed](0.0,0.0)(4.0,0.0)
 \psline[linecolor=gray,linewidth=0.5pt,linestyle=dashed](0.0,3.0)(4.0,3.0)
 \psline[linecolor=gray,linewidth=0.5pt,arrowsize=5pt]{->}(-3.0,3.0)(-2.5,2.5) 
 \psline[linecolor=gray,linewidth=0.5pt,arrowsize=5pt]{<-}(-2.5,2.5)(-2.0,2.0)
 \psline[linecolor=gray,linewidth=0.5pt,arrowsize=5pt]{<-}(-3.0,2.0)(-2.5,2.5)
 \psline[linecolor=gray,linewidth=0.5pt,arrowsize=5pt]{->}(-2.5,2.5)(-2.0,3.0)
 \psellipticarc[linecolor=blue,linewidth=2.0pt]{-}(-3.0,2.5)(0.353553,0.353553){315}{45}
 \psellipticarc[linecolor=blue,linewidth=2.0pt]{-}(-2.0,2.5)(0.353553,0.353553){135}{225}
 \rput[Bc](-2.5,1.8){$1$}
 \psline[linecolor=gray,linewidth=0.5pt,arrowsize=5pt]{->}(-1.5,3.0)(-1.0,2.5) 
 \psline[linecolor=gray,linewidth=0.5pt,arrowsize=5pt]{<-}(-1.0,2.5)(-0.5,2.0)
 \psline[linecolor=gray,linewidth=0.5pt,arrowsize=5pt]{<-}(-1.5,2.0)(-1.0,2.5)
 \psline[linecolor=gray,linewidth=0.5pt,arrowsize=5pt]{->}(-1.0,2.5)(-0.5,3.0)
 \psellipticarc[linecolor=blue,linewidth=2.0pt]{-}(-1.0,3.0)(0.353553,0.353553){225}{315}
 \psellipticarc[linecolor=blue,linewidth=2.0pt]{-}(-1.0,2.0)(0.353553,0.353553){45}{135}
 \rput[Bc](-1.0,1.8){$x$}
 \psline[linecolor=gray,linewidth=0.5pt,arrowsize=5pt]{<-}(-3.0,1.5)(-2.5,1.0) 
 \psline[linecolor=gray,linewidth=0.5pt,arrowsize=5pt]{->}(-2.5,1.0)(-2.0,0.5)
 \psline[linecolor=gray,linewidth=0.5pt,arrowsize=5pt]{->}(-3.0,0.5)(-2.5,1.0)
 \psline[linecolor=gray,linewidth=0.5pt,arrowsize=5pt]{<-}(-2.5,1.0)(-2.0,1.5)
 \psellipticarc[linecolor=blue,linewidth=2.0pt]{-}(-3.0,1.0)(0.353553,0.353553){315}{45}
 \psellipticarc[linecolor=blue,linewidth=2.0pt]{-}(-2.0,1.0)(0.353553,0.353553){135}{225}
 \rput[Bc](-2.5,0.3){$1$}
 \psline[linecolor=gray,linewidth=0.5pt,arrowsize=5pt]{<-}(-1.5,1.5)(-1.0,1.0) 
 \psline[linecolor=gray,linewidth=0.5pt,arrowsize=5pt]{->}(-1.0,1.0)(-0.5,0.5)
 \psline[linecolor=gray,linewidth=0.5pt,arrowsize=5pt]{->}(-1.5,0.5)(-1.0,1.0)
 \psline[linecolor=gray,linewidth=0.5pt,arrowsize=5pt]{<-}(-1.0,1.0)(-0.5,1.5)
 \psellipticarc[linecolor=blue,linewidth=2.0pt]{-}(-1.0,1.5)(0.353553,0.353553){225}{315}
 \psellipticarc[linecolor=blue,linewidth=2.0pt]{-}(-1.0,0.5)(0.353553,0.353553){45}{135}
 \rput[Bc](-1.0,0.3){$x^{-1}$}
\end{pspicture}
$$
\caption{Vertices, weights and sample configuration for
  dense polymers on a square lattice of width $L=3$. Boundary
  conditions are free in the horizontal (space) direction and periodic
  in the vertical (imaginary time) direction. The alternating
  $\square,\bar{\square}$ representations correspond to
  a lattice orientation, conserved along each loop.}
   \label{fig1}
\end{figure}

\paragraph{SUSY formalism for dense polymers.} The universality class
of dense polymers is generically obtained when one forces a finite
number of self-avoiding loops or walks to fill up a fraction of space
$\rho > 0$. When $\rho=1$ for finite systems, one obtains Hamiltonian
walks, the polymer limit of fully-packed loop models.  For $\rho=1$
the CFT is lattice dependent, i.e., universality breaks down
\cite{JK}.  Nevertheless, the loop model in Fig.~\ref{fig1}
is in the generic dense polymer universality class.

The CFT of dense polymers has $c=-2$; we review it below.  It was
discovered a few years back \cite{JRS} that if one allows four-leg
crossings the model flows to a different universality class with
$c=-1$, and trivial geometrical exponents. Such crossings imply that
loops no longer conserve the lattice orientation of Fig.~\ref{fig1},
indicating that a crucial symmetry is broken.

To identify this symmetry we need to get into a bit of algebra. We
consider the lattice in Fig.~\ref{fig1} and a transfer matrix $T$
propagating vertically.
We introduce a supersymmetric (SUSY) formulation \cite{grl,RS}: each
edge carries a ${\mathbb Z}_2$-graded vector space of dimensions $m+n$
(resp.\ $n$) for the even bosonic (resp.\ odd fermionic) subspace
($m+n,n\geq 0$ are integers).  We label edges $i=0,1,\ldots,2L-1$ for a
system of width $2L$.  The ${\mathbb Z}_2$ space is chosen as the
fundamental $\square$ of the Lie superalgebra gl($m+n|n$) 
for $i$ even
(down arrow), and its dual $\bar{\square}$ for $i$ odd (up arrow).
$T$ acts on the graded tensor product ${\cal
  H}=\left(\square\otimes\bar{\square}\right)^{\otimes L}$ (this
``Hilbert'' space has in fact an indefinite inner product).

To construct $T$ for critical dense polymers we first observe that,
for generic $m$, the tensor products $\square\otimes \bar{\square}$
and $\bar{\square}\otimes\square$ decompose as the direct sum of the
singlet and the adjoint. The projectors on the singlet obey the
Temperley-Lieb algebra relations $E_i^2=mE_i$,
$\left[E_i,E_j\right]=0$ for $|i-j|>2$, and $E_iE_{i\pm 1}E_i=E_i$
(and here $m=0$). The $E_i$ can be expressed as quadratic terms in the
SUSY generators, and are closely related with the Casimir
\cite{RS}. We have $T\equiv T_1 T_3\cdots T_{2L-3}T_0T_2\cdots
T_{2L-2}$, where $T_i=1+xE_i$.
By taking either of the two terms in $T_i$ for each vertex, the
expansion of Fig.~\ref{fig1} is obtained, with a power of $x$ for each
vertex, and a factor $(n+m)-n={\rm str}\, 1=m$ for each loop.
The latter equals the supertrace in the fundamental representation
(denoted $\rm str$) of $1$, since states in ${\cal H}$ flow around the
loop. This holds whether the loop be topologically nontrivial or
homotopic to a point. Isotropic dense polymers now correspond to $m=0$
and $x=1$. Note that when $m=0$, the tensor product $\square\otimes \bar{\square}$ 
is indecomposable; $E_i$  can then be defined as  the unique invariant coupling (on two sites) other than the identity. The rest of the discussion is unchanged.

Letting $x\rightarrow 0$ allows one to extract the spin chain
hamiltonian $H\propto -\sum_i E_i$ acting on ${\cal H}$; the scale of
$H$ is chosen to ensure conformal invariance. The interaction is
simply the invariant quadratic coupling (Casimir), providing a natural
generalization of the Heisenberg chain to the $gl(n+m|n)$ case.

For such models, there is a corresponding continuum quantum field
theory \cite{RS}, which is a nonlinear $\sigma$-model with target space the
symmetry supergroup [here U($n+m|n$)], modulo the isotropy supergroup
of the highest weight state (see \cite{affleck,rs1} for related
non-SUSY examples, and \cite{z,grl} for SUSY random
fermion problems). Here we obtain
$U(n+m|n)/U(1)\times U(n+m-1|n) \cong {\bf CP}^{n+m-1|n}$, a
SUSY version of complex projective space. Moreover, the
 mapping shows that this model has a topological angle
$\theta=\pi$.

\paragraph{Dense polymers and sigma models.} Let us now make things
concrete: the fields can be represented by complex components $z^a$
($a=1,\ldots,n+m$) and $\zeta^\alpha$ ($\alpha=1,\ldots,n$), where
$z^a$ is commuting, $\zeta^\alpha$ is anticommuting. In these
coordinates, at each point in spacetime, the solutions to the
constraint $z_a^\dagger z^a+\zeta_\alpha^\dagger \zeta^\alpha=1$ (we
use the conjugation $\dagger$ that obeys
$(\eta\xi)^\dagger=\xi^\dagger\eta^\dagger$ for any $\eta$, $\xi$),
modulo $U(1)$ phase transformations $z^a\mapsto e^{iB}z^a$,
$\zeta^\alpha\mapsto e^{iB}\zeta^\alpha$, parametrize ${\bf
  CP}^{n+m-1|n}$. The Lagrangian density in 2D Euclidean spacetime is
\begin{eqnarray}
{\cal L}&=&\frac{1}{2g_\sigma^2}\left[(\partial_\mu
-ia_\mu)z_a^\dagger (\partial_\mu+ia_\mu)z^a
\right. \label{cpnlag} \\
&+& \left. (\partial_\mu-ia_\mu)\zeta_\alpha^\dagger
(\partial_\mu+ia_\mu)\zeta^\alpha\right]
+\frac{i\theta}{2\pi}(\partial_\mu a_\nu-\partial_\nu
a_\mu), \nonumber
\end{eqnarray}
where
$a_\mu=\frac{i}{2}[z_a^\dagger\partial_\mu z^a+\zeta_\alpha
^\dagger\partial_\mu \zeta^\alpha-(\partial
z_a^\dagger)z^a-(\partial \zeta_\alpha^\dagger)\zeta^\alpha]$
for $\mu=1,2$.  The fields are subject to the constraint, and under
the $U(1)$ gauge invariance $a_\mu$ transforms as a gauge potential; a
gauge must be fixed in any calculation. This set-up is similar to the
non-SUSY ${\bf CP}^{m-1}$ model in \cite{dadda,wit}. The coupling
constants are $g_\sigma^2$, the usual $\sigma$-model coupling (there
is only one such coupling, because the target supermanifold is a supersymmetric
space, and hence the metric on the target space is unique up to a
constant factor), and $\theta$, the coefficient of the topological
term ($\theta$ is defined modulo $2\pi$).

First we note a well-known important point about the SUSY models: the
physics is the same for all $n$, in the following sense. For example,
in the present model, correlation functions of operators that are
local functions (possibly including derivatives) of components $a\leq
n_1+m$, $\alpha\leq n_1$ for some $n_1$ are equal for any $n \ge n_1$,
due to cancellation of the ``unused'' even and odd index values. This
can be seen in perturbation theory because the unused index values
appear only in summations over closed loops, and their contributions
cancel, but is also true nonperturbatively (it can be shown in the
lattice constructions we discuss below). In particular, the
renormalization group (RG) flow of the coupling $g_\sigma^2$ is the
same as for $n=0$, a non-SUSY $\sigma$-model. For the case of
${\bf CP}^{n+m-1|n}$, the perturbative $\beta$-function is the same as
for ${\bf CP}^{m-1}$, namely (we will not be precise about the
normalization of $g_\sigma^2$)
\begin{equation}
 \frac{{\rm d}g_\sigma^2}{{\rm d}\xi}=
 \beta(g_\sigma^2)=mg_\sigma^4+O(g_\sigma^6) \,,
\label{umbet}
\end{equation}
where $\xi=\log L$, with $L$ the length scale at which the
coupling is defined [see e.g.\ \cite{wegner}, eq.\ (3.4)].
(The $\beta$-function for $\theta$ is zero in perturbation theory,
and that for $g_\sigma^2$ is independent of $\theta$.) For $m>0$,
if the coupling is weak at short length scales, then it flows to
larger values at larger length scales. For $\theta\neq\pi$ (mod
$2\pi$), the coupling becomes large, the U($n+m|n$) symmetry is
restored, and the theory is massive. However, a transition is
expected at $\theta=\pi$ (mod $2\pi$). For $m>2$, this transition
is believed to be first order, while it is second order for $m\leq
2$ \cite{affleck}. In the latter case, the system with
$\theta=\pi$ flows to a conformally-invariant fixed-point theory.
At the fixed point, a change in $\theta$ is a relevant
perturbation that makes the theory massive. 

For $m=0$, the perturbative $\beta$-function vanishes
identically. This can be seen either from direct calculations, which
have been done to at least four-loop order \cite{wegner}, or from an
argument similar to that in \cite{berk}: for $n=1$, the $\sigma$-model
reduces to the massless free fermion theory \cite{Kausch}
${\cal L}\propto \frac{1}{2g_\sigma^2}\partial_\mu
\zeta^\dagger\partial_\mu\zeta$
and further the $\theta$-term becomes trivial in this case. Thus, for
all $\sigma$-model couplings $g_\sigma^2>0$, the $n=1$ theory is
non-interacting. The free-fermion theory is conformal with $c=-2$, and
$\theta$ is a redundant perturbation, as it does not appear in the
action (a similar argument appeared in Ref.\ \cite{bsz}). By the above
argument, conformal invariance with $c=-2$ should hold for all $n$,
and also for all $g_\sigma^2$ and $\theta$, though the action is no
longer non-interacting in general. Thus the $\beta$-function also
vanishes non-perturbatively. In general, the scaling dimensions will
vary with the coupling $g_\sigma^2$, so changing $g_\sigma^2$ is an
exactly marginal perturbation, though for $n=1$ the coupling can be
scaled away, so there is no dependence on the coupling in the
exponents related to those multiplets of operators that survive at
$n=1$. Hence for $n=1$, the exactly-marginal perturbation that changes
$g_\sigma^2$ is redundant.

\paragraph{Introducing the six-leg crossings.}  For $n=1$, the
$\sigma$-model thus does not exhibit very interesting physics. It also
describes very few observables in the dense polymer problem. Indeed,
the underlying algebra psl($1|1$) does not admit any non-trivial
invariant tensor, so the only $\ell$-leg  operators
present have $\ell=0,2$, and they are moreover degenerate---and part
of an indecomposable block. These observables are expected to be
present in all theories with $n>1$ as well, and to not depend on the
coupling constant $g_\sigma^2$. However, for $n>1$, more observables
are possible. E.g.\  $\ell$-leg operators for all even $\ell$ exist, and
correspond to fully symmetric invariant tensors of psl($n|n$); there
is no reason why the corresponding conformal dimensions should not
depend on $g_\sigma^2$, and indeed we will shortly see that they do.

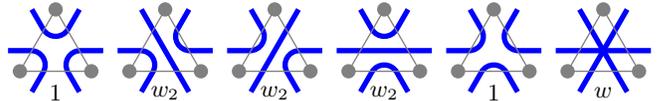
\begin{figure}
$$
\psset{unit=0.95cm}
\begin{pspicture}(-0.1,0.1)(1.1,0.966)
 \psline[linearc=0.2,linecolor=blue,linewidth=2.0pt](-0.166,0.289)(0.5,0.289)(0.167,-0.289)
 \psline[linearc=0.2,linecolor=blue,linewidth=2.0pt](0.167,0.866)(0.5,0.289)(0.833,0.866)
 \psline[linearc=0.2,linecolor=blue,linewidth=2.0pt](1.166,0.289)(0.5,0.289)(0.833,-0.289)
 \psline[linecolor=gray,linewidth=0.5pt](0.0,0.0)(1.0,0.0)(0.5,0.866)(0.0,0.0)
 \pscircle*[linecolor=gray](0.0,0.0){0.1}
 \pscircle*[linecolor=gray](1.0,0.0){0.1}
 \pscircle*[linecolor=gray](0.5,0.866){0.1}
 \rput[Bc](0.5,-0.3){$1$}
\end{pspicture} \quad
\begin{pspicture}(-0.1,0.1)(1.1,0.966)
 \psline[linearc=0.2,linecolor=blue,linewidth=2.0pt](1.166,0.289)(0.5,0.289)(0.833,0.866)
 \psline[linearc=0.2,linecolor=blue,linewidth=2.0pt](0.167,0.866)(0.833,-0.289)
 \psline[linearc=0.2,linecolor=blue,linewidth=2.0pt](0.167,-0.289)(0.5,0.289)(-0.166,0.289)
 \psline[linecolor=gray,linewidth=0.5pt](0.0,0.0)(1.0,0.0)(0.5,0.866)(0.0,0.0)
 \pscircle*[linecolor=gray](0.0,0.0){0.1}
 \pscircle*[linecolor=gray](1.0,0.0){0.1}
 \pscircle*[linecolor=gray](0.5,0.866){0.1}
 \rput[Bc](0.5,-0.3){$w_2$}
\end{pspicture} \quad
\begin{pspicture}(-0.1,0.1)(1.1,0.966)
 \psline[linearc=0.2,linecolor=blue,linewidth=2.0pt](-0.166,0.289)(0.5,0.289)(0.167,0.866)
 \psline[linearc=0.2,linecolor=blue,linewidth=2.0pt](0.167,-0.289)(0.833,0.866)
 \psline[linearc=0.2,linecolor=blue,linewidth=2.0pt](0.833,-0.289)(0.5,0.289)(1.166,0.289)
 \psline[linecolor=gray,linewidth=0.5pt](0.0,0.0)(1.0,0.0)(0.5,0.866)(0.0,0.0)
 \pscircle*[linecolor=gray](0.0,0.0){0.1}
 \pscircle*[linecolor=gray](1.0,0.0){0.1}
 \pscircle*[linecolor=gray](0.5,0.866){0.1}
 \rput[Bc](0.5,-0.3){$w_2$}
\end{pspicture} \quad
\begin{pspicture}(-0.1,0.1)(1.1,0.966)
 \psline[linearc=0.2,linecolor=blue,linewidth=2.0pt](0.167,0.866)(0.5,0.289)(0.833,0.866)
 \psline[linearc=0.2,linecolor=blue,linewidth=2.0pt](-0.166,0.289)(1.166,0.289)
 \psline[linearc=0.2,linecolor=blue,linewidth=2.0pt](0.167,-0.289)(0.5,0.289)(0.833,-0.289)
 \psline[linecolor=gray,linewidth=0.5pt](0.0,0.0)(1.0,0.0)(0.5,0.866)(0.0,0.0)
 \pscircle*[linecolor=gray](0.0,0.0){0.1}
 \pscircle*[linecolor=gray](1.0,0.0){0.1}
 \pscircle*[linecolor=gray](0.5,0.866){0.1}
 \rput[Bc](0.5,-0.3){$w_2$}
\end{pspicture} \quad
\begin{pspicture}(-0.1,0.1)(1.1,0.966)
 \psline[linearc=0.2,linecolor=blue,linewidth=2.0pt](0.167,-0.289)(0.5,0.289)(0.833,-0.289)
 \psline[linearc=0.2,linecolor=blue,linewidth=2.0pt](0.833,0.866)(0.5,0.289)(1.166,0.289)
 \psline[linearc=0.2,linecolor=blue,linewidth=2.0pt](-0.166,0.289)(0.5,0.289)(0.167,0.866)
 \psline[linecolor=gray,linewidth=0.5pt](0.0,0.0)(1.0,0.0)(0.5,0.866)(0.0,0.0)
 \pscircle*[linecolor=gray](0.0,0.0){0.1}
 \pscircle*[linecolor=gray](1.0,0.0){0.1}
 \pscircle*[linecolor=gray](0.5,0.866){0.1}
 \rput[Bc](0.5,-0.3){$1$}
\end{pspicture} \quad
\begin{pspicture}(-0.1,0.1)(1.1,0.966)
 \psline[linearc=0.2,linecolor=blue,linewidth=2.0pt](-0.166,0.289)(1.166,0.289)
 \psline[linearc=0.2,linecolor=blue,linewidth=2.0pt](0.167,-0.289)(0.833,0.866)
 \psline[linearc=0.2,linecolor=blue,linewidth=2.0pt](0.833,-0.289)(0.167,0.866)
 \psline[linecolor=gray,linewidth=0.5pt](0.0,0.0)(1.0,0.0)(0.5,0.866)(0.0,0.0)
 \pscircle*[linecolor=gray](0.0,0.0){0.1}
 \pscircle*[linecolor=gray](1.0,0.0){0.1}
 \pscircle*[linecolor=gray](0.5,0.866){0.1}
 \rput[Bc](0.5,-0.3){$w$}
\end{pspicture}
\psset{unit=1.0cm}
$$
  \caption{Vertices and weights for dense polymers on the triangular lattice.
  When $w=0$ this is equivalent \cite{WuLin} to a Potts model with spins on
  the circles and arbitrary interactions within the gray triangles.}
  \label{fig2}
\end{figure}

For this, we need to be able to tune $g_\sigma^2$ in the lattice
model. We propose doing so by allowing not four-leg but {\em six-leg}
crossings. This can be described most conveniently by going to the
hamiltonian formalism, and adding interactions that preserve the
symmetry. Four-leg crossings would then translate into a perturbation
of the type $P_{i,i+1}$ which exchanges spaces at position $i$ and
$i+1$. Since by construction our chain has alternating representations
this is not possible within gl($n+m|n$) symmetry, so forcing such
crossings breaks the symmetry down to the orthosymplectic subgroup.
On the other hand, six-leg crossings correspond to exchanging
representations at position $i,i+2$ while the one at $i+1$ just goes
through, and is perfectly compatible with the  $gl(n+m|n)$ symmetry (notice however that it breaks the extended symmetry discussed in \cite{RS2}). The
hamiltonian then becomes
\begin{equation}
H\propto -\sum_i (E_i+wP_{i,i+2}) \,. \label{pertham}
\end{equation}
Our first claim is that the continuum limit of (\ref{pertham}) is
described by the superprojective $\sigma$-model ${\bf CP}^{n-1|n}$ with
coupling $g_\sigma^2(w)$, at $\theta=\pi$.  Note that we could more
generally study the spectrum of the hamiltonian $ H\propto -\sum_i
\left[E_i+wP_{i,i+2}+w_2\left(E_iE_{i+1}+E_{i+1}E_i\right)\right]$.  The
symmetries are unchanged, and one expects the continuum limit to be
described by the same $\sigma$-model, with now $g_\sigma^2(w,w_2)$.
This is confirmed by numerical calculations. Finally, a more pleasant
realization of the same physics is provided by a model of dense
polymers on the triangular lattice, where six-leg crossings can
naturally take place; see Fig.~\ref{fig2}. We shall call the Boltzmann
weight of these vertices $w$ as well, and the same conclusions will
hold for this model as for the spin chain (\ref{pertham}).

We first check what happens for $n=1$, where everything can be
reformulated in terms of free fermion operators and their adjoints  \cite{RS2}
$f_i,f_i^\dagger$, obeying $\{f_i,f_{i'}\}=0$, $\{f_i,f_{i'}^\dagger\}=(-1)^i\delta_{i'}$
through
$E_i=(f^\dagger_i+f^\dagger_{i+1})(f_i+f_{i+1})$
and
$P_{i,i+2}=(-1)^i+(f_{i-1}^\dagger -f^\dagger_{i+1})(f_{i-1}-f_{i+1})$.
Since both are quadratic it is easy to show that the continuum limit
of (\ref{pertham}) is unchanged, with $w$ only affecting the sound
velocity and the fine structure of the Jordan blocks.

One can easily argue that the ground state energy is the same for the
$n=1$ and $n>1$ models, whence $c=-2$ independently of
$w$. This is confirmed by transfer matrix calculations for the model
in Fig.~\ref{fig2}.  The $\ell=2$ exponent is conjugate to the fractal
dimension of the loop, hence zero.

Numerical study of the $\ell>2$ leg exponents then clearly shows that
they are non-trivial, decreasing functions of $w$. To discuss this some
more we place ourselves in the simplest case of free boundary
conditions. The exponents at the special point $w=0$ are well known to
be
$h_\ell^0=h_{1,1+\ell}= \frac{\ell(\ell-2)}{8}$.
We next assume that $w\rightarrow\infty$ corresponds to the
weak-coupling limit of the $\sigma$-model, $g_\sigma^2\rightarrow
0$. This is qualitatively very reasonable: in the limit of large $w$,
the system almost splits into two subsystems with gl($n|n$) symmetry
involving only the fundamental or only its dual, with in both cases a
simple interaction of the type $P_{i,i+1}$. Such models are well-known
to be integrable, and their physics to be described by a weak-coupling
limit not unlike the  XXX ferromagnetic
spin chain. In such a limit, we can analyze the spectrum using the
minisuperspace approach, that is, by analyzing quantum mechanics on
the target manifold.  The spectrum of the Laplacian on the ordinary
projective space ${\bf CP}^{m-1}=U(m)/U(1) \times U(m-1)$ is
well-known to be of the form $E_l\propto 4l(l+m-1)$, so, setting
$m=0$, we find that \cite{ZhangZou} $h_l^{\rm
  wc}=g_\sigma^2{l(l-1)\over 2}$.
Here $l$ in an integer, which we can identify using psl($n|n$)
representation theory with $\ell/2$. Remarkably, $h_l^{\rm wc}$
coincides with the known result $h_\ell^0$ at $w=0$ (ordinary dense
polymers) if we identify $g_\sigma^2=1$ in that case.

\paragraph{Conjecture for the exact exponents.} 
We conjecture that  the boundary 
conformal dimensions in our model  are simply
linear in the Casimir of the associated representation of psl($n|n$).
This is  due  to the  structure of the
perturbation theory where the vanishing of the dual Coxeter number---the Casimir in the
adjoint--- suggests  exactness of  the minisuperspace approximation (see \cite{QSC} for a related case).  A more thorough 
study of this perturbation theory, together with non perturbative arguments, will appear elsewhere \cite{Draft}. For now, we simply propose that the exponents be given by 
\begin{equation}
 h_\ell = g_\sigma^2 \frac{\ell(\ell-2)}{8} \,,
 \label{conj}
\end{equation}
where $g_\sigma^2$ is a decreasing function of $w$, equal to unity
when $w=0$, and vanishing at large $w$.

\begin{figure}
  \vspace{-1.0cm}
  \centerline{\includegraphics[scale=0.35,angle=270]{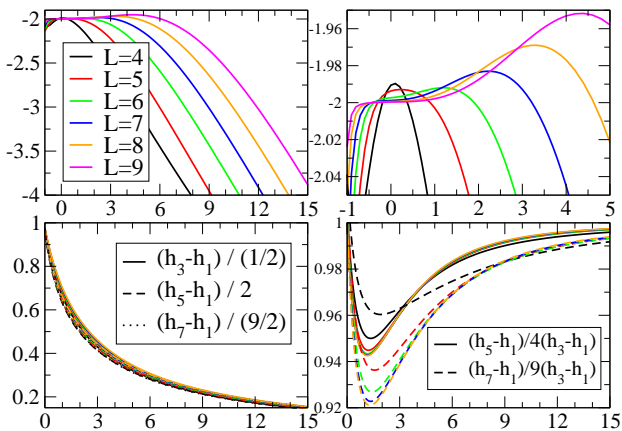}}
  \vspace{-0.7cm}
  \caption{The two upper panels show the central charge as a function
    of the intersection weight $w$ for width $L$ strips of the
    triangular lattice. In the text we show analytically that $c=-2$
    always. The lower left panel represents the effective
    coupling constant $g_\sigma^2$ extracted from (\ref{conj}) using
    different values of $\ell$. The collapse on a single curve is quite
    striking. The lower right panel shows details of the exponents, in
    particular the region close to $w=0$ where convergence appears
    actually less good than on the previous curves. }
  \label{fig4}
\end{figure}

This conjecture is compared with the results of exact diagonalizations
in Fig.~\ref{fig4} (lower panels), where we have represented the
function $g_\sigma^2(w)$ as extracted from (\ref{conj}) and various
$\ell$.  The different estimates collapse on a single curve over the
whole range of $w$ values, in agreement with the conjecture.

For the model of Fig.~\ref{fig2} it is technically difficult to study
operators with $\ell$ even. The $\sigma$-model formalism can however
be extended to $\ell$ odd, and the arguments leading to (\ref{conj}) extended to this case \cite{Draft}. Exact
diagonalization of the spin chain hamiltonian on $2L=18$ sites yields
results for $\ell$ even which look like the lower left panel of
Fig.~\ref{fig4}, except that $w$ now has a different meaning. The
sound velocity is determined from analytical results for the
$n=1$ case.

While the $\ell$-leg exponents for the usual dense polymers ($w=0$) agree with the general conjecture for $g_\sigma^2=1$, the fine structure of the spectrum at that point \cite{RS3} differs from the one of the sigma model. The situation seems similar to the one encountered in \cite{CS} for the supersphere sigma model, where the point $w=0$ is in fact singular. 

A related question concerns  periodic boundary conditions. In this case, the known
values of the {\em bulk} polymer exponents at $w=0$ are
$h_\ell = \frac{\ell^2-4}{32} = \frac{l^2-1}{8}$.
The fact that $h_6 =1$ provides an independent argument for the 
marginality of the $w$ perturbation. Meanwhile, note that $h_\ell$ now do
{\em not} have the minisuperspace form. For large $w$, one can however
argue that the minisuperspace form remains valid, as is confirmed
numerically. This suggests again that the point $w=0$ is singular. It could also be 
 that in the periodic case, the
arguments that the minisuperspace should be exact for any $w$ fail, which agrees
with the expectations  for a related model in \cite{QSC}. More work is needed to clarify
this point.

We checked that, within numerical accuracy, staggering the chain
produces similar results but with a coupling constant that now depends
on $w$ and the staggering parameter---that is, the $\theta$ angle in
the continuum limit. We e also studied  the effect of
coupling additional $\square$ or $\bar{\square}$ representations on
the boundary, which can be interpreted in terms of boundary $\theta$
angle \cite{Draft}. All the results are compatible with the
$\sigma$-model picture.

In conclusion, we have shown that allowing intersections where three lines cross 
profoundly modifies the dense polymer problem. It gives rise to a critical line of conformal field theories, with central charge $c=-2$, which  can be identified with 
 the long distance
limit of a conformal sigma model such as those studied in the AdS/CFT correspondence . Our identification leads moreover  to the
proposal of an exact formula (\ref{conj}) for the $\ell$-leg polymer
exponents in the boundary case, and opens the way to tackling the
sigma model using lattice techniques.

\paragraph{Acknowledgments:} JLJ and HS were supported by the ANR, and HS
by the ESF Network INSTANS. NR was supported by NSF grant no.\
DMR-0706195.


\end{document}